\documentclass[twocolumn,nopacs,floatfix,amsmath,nofootinbib,amssymb,preprintnumbers]{revtex4}
\usepackage{graphicx,color,dcolumn,booktabs,bm}
\usepackage{txfonts}
\usepackage{slashed}
\usepackage{epstopdf}
\usepackage{graphicx,color,dcolumn,booktabs,bm}
\usepackage[colorlinks,
            citecolor=blue,
            anchorcolor=red,
            menucolor=red,
            linkcolor=red,
            filecolor=red,
            runcolor=red,
            urlcolor=blue,
            frenchlinks=red]{hyperref}

\begin{document}
\title{Connection between near the $D_s^+D_s^-$ threshold enhancement in $B^+ \to D_s^+D_s^-K^+$ and conventional charmonium $\chi_{c0}(2P)$}
\author{Dan Guo$^{1,3}$}\email{guod13@lzu.edu.cn}
\author{Jun-Zhang Wang$^{2,1}$}\email{wangjzh2022@pku.edu.cn}
\author{Dian-Yong Chen$^{4,5}$}\email{chendy@seu.edu.cn}
\author{Xiang Liu$^{1,3,5}$}\email{xiangliu@lzu.edu.cn}

\affiliation{
$^1$School of Physical Science and Technology, Lanzhou University, Lanzhou 730000, China\\
$^2$School of Physics and Center of High Energy Physics, Peking University, Beijing 100871, China\\
$^3$Research Center for Hadron and CSR Physics, Lanzhou University and Institute of Modern Physics of CAS, Lanzhou 730000, China\\
$^4$School of Physics, Southeast University, Nanjing 210094, China\\
$^5$Lanzhou Center for Theoretical Physics, Key Laboratory of Theoretical Physics of Gansu Province and Frontier Science Center for Rare Isotopes, Lanzhou University, Lanzhou 730000, China
}

\begin{abstract}
Focusing on recent measurement of $B^+ \to D_s^+D_s^-K^+$ process given by the LHCb Collaboration,
we propose that this newly observed near  the $D_s^+D_s^-$ threshold enhancement can be due to the contribution of the $\chi_{c0}(2P)$, which is a $P$-wave charmonium below the $D_s^+D_s^-$ threshold. 
By performing a combined fit to the measured $D_s^+D_s^-$, $D_s^+K^+$, and $D_s^-K^+$ invariant mass spectra, introducing the $\chi_{c0}(2P)$ can well reproduce the near threshold enhancement in the $D_s^+D_s^-$ invariant mass spectrum. When depicting the whole $D_s^+D_s^-$ invariant mass spectrum, the contributions from
higher charmonium $\psi(4230)$ and charmoniumlike state $X_0(4140)$ are obvious. In addition, a charmed meson $D^*(3^3S_1)$ with mass around 3015 MeV is found to be important to
depict the $D_s^-K^+$ invariant mass spectrum well. 
Especially, the importance of node effect of the spatial wave function of the $\chi_{c0}(2P)$ is revealed, by which the anomaly of the ratio $\Gamma(X\to D^+D^-)/\Gamma(X\to D_s^+D_s^-)$ indicated by LHCb can be explained well. Finally, our scenario of the newly observed enhancement structure resulted from the $\chi_{c0}(2P)$ is enforced.

\end{abstract}

\maketitle
\section{Introduction}\label{sec1}

Very recently, the LHCb Collaboration found a novel phenomenon of near threshold
enhancement in the $D_s^+D_s^-$ invariant mass distribution of the $B^+\to D_s^+D_s^-K^+$ process \cite{LHCb:2022vsv}. If adopting the Breit-Wigner formula to depict this enhancement structure, the extracted resonant parameters are
$M=3956\pm 5({\rm stat.})\pm 10({\rm syst.})$ MeV and
$\Gamma=43\pm 13({\rm stat.})\pm 8({\rm syst.})$ MeV \cite{LHCb:2022vsv}. Thus,
this enhancement structure is refereed to be the $X(3960)$ before decoding its nature. 
In addition, the spin-parity test from LHCb suggests that the $X(3960)$ favors $J^{PC}=0^{++}$. When checking the data of the $D_s^+D_s^-$ invariant mass distribution, a dip around 4.14 GeV can be found, which was due to another $J^{PC}=0^{++}$ state $X_0(4140)$ by LHCb \cite{LHCb:2022vsv}.

For understanding this novel phenomenon, some theoretical explanations were proposed. 
The $X(3960)$ was explained as a threshold structure due to the $D_s\bar{D}_s$ interaction, where the interpretation of a virtual or bound state below the $D_s^+D_s^-$ threshold \cite{Ji:2022uie,Xie:2022lyw}, and the $D_s^+D_s^-$ molecule description from the QCD sum rules method \cite{Xin:2022bzt} and the one-boson-exchange  model \cite{Chen:2022dad} were proposed. In Ref. \cite{Bayar:2022dqa}, the near threshold enhancement signal of the $X(3960)$ was interpreted to be the off-shell contribution of the $\chi_{c0}(3930)$, which is a $P$-wave charmonium state found in the $B^+\to D^+D^-K^+$ decay \cite{LHCb:2020pxc}.

A main reason why most of theoretical groups tried to categorize the newly observed $X(3960)$ to be an exotic hadronic state is due to the measured ratio
\begin{eqnarray}\label{LHCbratio}
\frac{\Gamma(X\to D^+D^-)}{\Gamma(X\to D_s^+D_s^-)}&=&\frac{\mathcal{B}[B^+\to D^+D^-K^+]\mathcal{FF}^X_{B^+\to D^+D^-K^+}}{\mathcal{B}[B^+\to D_s^+D_s^-K^+]\mathcal{FF}^X_{B^+\to D_s^+D_s^-K^+}} \nonumber \\
&=&0.29\pm0.09\pm0.10\pm0.08,
\end{eqnarray}
given by LHCb when treating the $X(3960)$ in the $B^+\to D_s^+D_s^-K^+$ process \cite{LHCb:2022vsv} and the $\chi_{c0}(3930)$ in the $B^+\to D^+D^-K^+$ decay \cite{LHCb:2020pxc} as the same state (see Ref. \cite{LHCb:2022vsv} for more details). Usually, the $s\bar{s}$ quark pair excited from vacuum is harder  than the $u\bar{u}$ and $d\bar{d}$ pairs. The ratio is anomalous for a higher charmonium as indicated by LHCb \cite{LHCb:2022vsv}. 
It seems to provide enough motivation to assign the newly observed $X(3960)$ as an exotic hadronic state. However, we must face the fact that the coupled-channel effect is obvious for these $2P$ states of charmonium \cite{Duan:2020tsx}. We still should be cautious when definitely making a conclusion of the $X(3960)$ to be an exotic state.

Thus, we briefly introduce how the coupled-channel effect plays the role in the spectroscopy behavior of these $2P$ states of charmonium. When solving the low mass puzzle of the $X(3872)$ \cite{Godfrey:1985xj,Barnes:2005pb}, the importance of the coupled-channel was realized \cite{Barnes:2003vb,Kalashnikova:2005ui,Ortega:2009hj,Duan:2020tsx}, where the bare mass of the $\chi_{c1}(2P)$ can be decorated by the nearby $D\bar{D}^\ast $ channel. After observing the charmoniumlike state $X(3915)$ \cite{Belle:2009and} in the $\gamma\gamma\to J/\psi\omega$ fusion process, the Lanzhou group indicated that
the $X(3915)$ must have $0^{++}$ quantum number, which is a good candidate of the charmonium $\chi_{c0}(2P)$ \cite{Liu:2009fe}. Later, this prediction was
confirmed by the BaBar Collaboration \cite{BaBar:2012nxg}. Thus, the Particle Data Group (PDG) once collected the $X(3915)$ as the charmonium $\chi_{c0}(3915)$ into {\it The Review of Particle Physics (2012)} \cite{ParticleDataGroup:2012pjm} in a short time. 
Under this assignment to the $X(3915)$, we have to explain why
the mass gap between the $X(3915)$ and $Z(3930)\equiv\chi_{c2}(2P)$ is smaller than that between the $\chi_{c0}(1P)$ and $\chi_{c2}(1P)$, or that between the $\chi_{b0}(2P)$ and $\chi_{b2}(2P)$ \cite{Godfrey:1985xj,Barnes:2005pb,Guo:2012tv}. Finally, the Lanzhou group clarified this issue by introducing the coupled-channel effect, and found that the node effect of the spatial wave functions of the discussed $2P$ charmonoia is crucial to exactly reproduce the narrow mass gap between the $X(3915)$ and $Z(3930)$
\cite{Duan:2020tsx}, which supports
the assignment of the $X(3915)$ as a $\chi_{c0}(2P)$ charmonium. 
In 2020, LHCb observed the $\chi_{c0}(3930)$ and $\chi_{c2}(3930)$ simultaneously in the $B^+\to D^+D^-K^+$ decay \cite{LHCb:2020pxc}, which is the first signal of the $\chi_{c0}(2P)\to D\bar{D}$ decay mode. 
This observation confirms
that the $\chi_{c0}(2P)$ state has narrow width \cite{Liu:2009fe}, and also checks the prediction from the Lanzhou group that two $P$-wave charmonia $\chi_{c0}(2P)$ and $\chi_{c2}(2P)$ are close to each other \cite{Chen:2012wy,Duan:2020tsx}. 
By these efforts, the name $\chi_{c0}(3915)$ appeared in {\it The Review of Particle Physics (2022)} \cite{Workman:2022ynf} again.

Coming back to the LHCb's observation of the near threshold enhancement in the $D_s^+D_s^-$ invariant mass spectrum \cite{LHCb:2022vsv}, we should carefully check whether this near threshold enhancement can be resulted by the $\chi_{c0}(2P)$ under considering the unquenched effect. In this work, we carry out a serious study on this issue. According to the measured mass of the $\chi_{c0}(3930)$, we conclude that the $\chi_{c0}(2P)$ is below and near the $D_s^+D_s^-$ threshold. However, the $\chi_{c0}(2P)$ can still contribute to $B\to D_s^+D_s^- K^+$ and should have obvious effect to the near threshold enhancement in the $D_s^+D_s^-$ invariant mass spectrum if the width effect of the $\chi_{c0}(2P)$ is considered. For depicting the dip at 4.14 GeV in the $D_s^+D_s^-$ invariant mass spectrum, we still need to introduce a scalar $X_0(4140)$ state similar to the treatment of LHCb \cite{LHCb:2022vsv}. Besides, we notice event accumulation around 4.2 GeV, which also inspires us to test the contribution of the higher vector charmonium $\psi(4230)$ \cite{BESIII:2014rja,BESIII:2016bnd,BESIII:2016adj,Wang:2019mhs,Wang:2017sxq,Qian:2021neg,Chen:2014sra,Chen:2017uof} to $B\to D_s^+D_s^- K^+$. In fact, these intermediate states like the $\chi_{c0}(2P)$, $X_0(4140)$, and $\psi(4230)$ have the reflections in the $D_s^- K^+$, and $D_s^+K^+$ invariant mass spectra.  However, if checking the $D_s^- K^+$ invariant mass spectrum, one may find the event cluster around 3 GeV compared with the $D_s^+ K^+$ invariant mass distributions, which indicates the contributions from the higher charmed mesons~\cite{Song:2015fha,Wang:2016krl}. Thus, in the present work, we perform a combined fit to these measured three invariant mass distributions of the  $B\to D_s^+D_s^- K^+$ decay by including the charmonia/charmonium-like intermediate states and charmed mesons intermediate states simultaneously to test our proposal. 
Moreover, if making a connection between the near threshold enhancement and the $\chi_{c0}(2P)$, we still need to suitably explain the anomaly of the ratio shown in Eq. (\ref{LHCbratio}). Inspired by the node effect to solve the small mass gap between the $X(3915)$ and $Z(3930)$ \cite{Duan:2020tsx}, we introduce the node effect, and reproduce the anomalous ratio in Eq. (\ref{LHCbratio}). In this work, we want to emphasize that this anomaly of the ratio of $\Gamma(X\to D^+D^-)/\Gamma(X\to D_s^+D_s^-)$ does not conflict with the $P$-wave charmonium assignment to depict the near threshold enhancement in the $D_s^+D_s^-$ invariant mass spectrum. Instead, reproducing this anomalous ratio in the present work may further enforce our scenario. 

This work is organized as follows. In the following Section, we focus on the interpretation of three invariant mass distributions, and in Sec. \ref{sec3}, we present an explanation to the abnormal coupling ratio. Finally, this work ends with a short summary in Sec. \ref{sec4}.

\section{Decoding the enhancement structures in the invariant mass spectra of $B^+\to D_s^+D_s^-K^+$ }\label{sec2}

\begin{figure}[htbp]
    \centering
\includegraphics[width=0.48\textwidth]{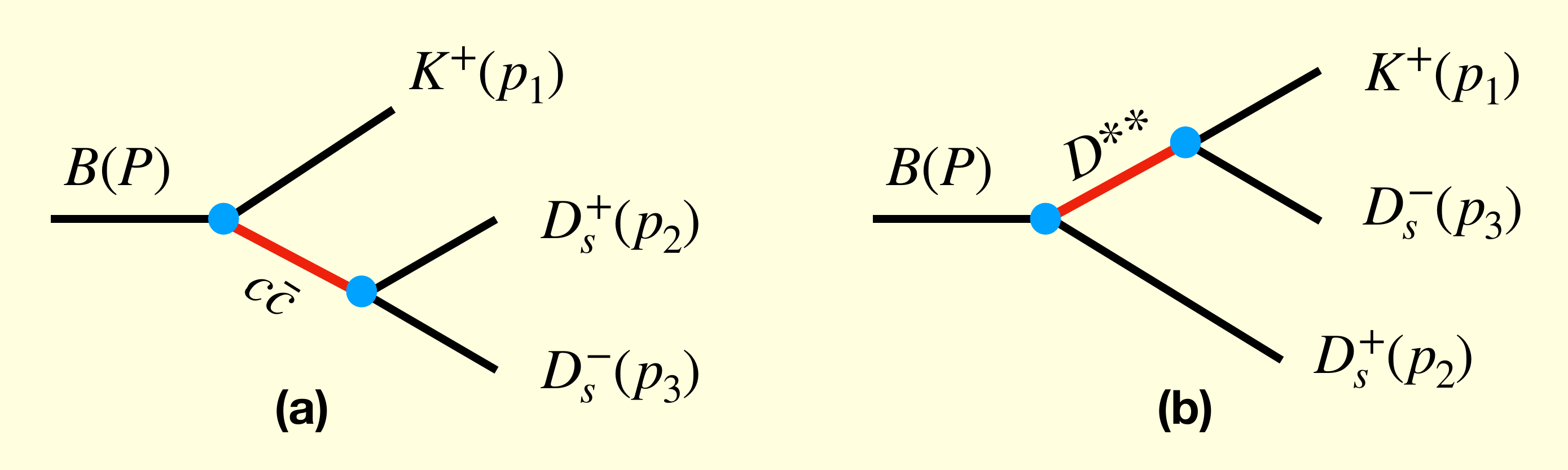}
    \caption{The schematic diagrams of different intermediate resonant contributions to the $B^+(P)\to K^+(p_1)D_s^+(p_2)D_s^-(p_3)$ decay. Diagram (a) presents the intermediate charmonium contribution, while diagram (b) stands for the charmed meson contribution.}
    \label{feynman}
\end{figure}

As discussed in the Introduction, there are two different kinds of intermediate state contributions to the process $B^+ \to K^+ D_s^- D_s^+$ as shown in Fig.~\ref{feynman}. For the processes corresponding to Fig.~\ref{feynman} (a), we consider the contributions from charmonia $\chi_{c0}(2P)$, $\psi(4230)$, and charmoniumlike $X(4140)$. For the process shown in Fig.~\ref{feynman} (b), all the possible charmed mesons above the $D_s^+K^-$ threshold with proper $J^P$ quantum numbers like $0^+$, $1^-$ and $2^+$ could be as the intermediate state. According to the spectroscopy behavior of the higher charmed mesons~\cite{Song:2015fha,Wang:2016krl}, there exist several candidates like the $D_1^\ast (2600)$, $D_2^\ast (2460)$, $D^\ast (2^3P_0)$, $D^\ast (1^3D_1)$, $D^\ast (3^3S_1)$, $D^\ast (2^3P_2)$ and $D^\ast (1^3F_2)$~\cite{Song:2015fha,Workman:2022ynf}. For depicting the event cluster around 3 GeV shown in the $D_s^-K^+$ invariant mass spectrum, finally we consider the charmed mesons with mass around 3 GeV for simplicity. Since the widths of the $D^\ast (2^3P_0)$, $D^\ast (1^3D_1)$ and $D^\ast (1^3F_2)$ are all larger than 200 MeV~\cite{Song:2015fha,Workman:2022ynf}, 
only the $D^\ast (3^3S_1)$ with the width to be 80.36 MeV is promising when refining the $D_s^-K^+$, $D_s^+K^+$ and $D_s^+D_s^-$ mass spectra simultaneously.

As discussed above, two kinds of intermediate resonant contributions are considered for the $B^+\to K^+D_s^+D_s^-$ process, which are $B^+ \to K^+ (c\bar{c}) \to K^+ [(c\bar{c}) \to D_s^+ D_s^-]$ and $B\to \bar{D}^{\ast0}(3^3S_1) D_s^+ \to  (\bar{D}^{\ast0} (3^3S_1)\to D_s^- K^+) D_s^+$. In this work, the effective Lagrangian approach is employed to describe the $B^+\to D_s^+D_s^-K^+$ decay, which has been successfully applied to describe other similar $B$ decay processes \cite{Duan:2021bna,Wang:2021crr}. Since the primary concern of this work is to explore the role of different intermediate states in describing the line shape of invariant mass distribution of $B^+\to D_s^+D_s^-K^+$, the effective Lagrangians relevant to the weak decay of $B$ meson can be constructed by considering the invariance under the isospin and parity transformation, which can be written as
\begin{eqnarray}
  \mathcal{L}_{B X K} &=& ig_{B X K}X B \bar{K}, \nonumber\\
  \mathcal{L}_{B\psi K} &=& ig_{B \psi K} \psi_\mu(\partial^\mu B^{\dag}K-B\partial^\mu K^{\dag}), \\
  \mathcal{L}_{B D^\ast D_s} &=& ig_{B D^\ast D_s} D^\ast _\mu(\partial^\mu D_s^\dag B-D_s\partial^\mu B^\dag).\nonumber
\end{eqnarray}
Here, $X$ refers to the scalar $\chi_{c0}(2P)$ and $X_0(4140)$, while
the $\psi(4230)$ and $D^\ast(3^3S_1)$ are abbreviated to be $\psi$ and $D^\ast$, respectively.

For the $(c\bar{c})\to D_s^+ D_s^-$ and $D^\ast(3^3S_1) \to DK$ interactions, the relevant effective Lagrangians based on the heavy quark symmetry and chiral symmetry are \cite{Duan:2021bna, Huang:2021kfm,Chen:2012nva,Wang:2020kej}
\begin{eqnarray}
  \mathcal{L}_{X D_s\bar{D}_s} &=& ig_{XD_s\bar{D}_s}XD_s\bar{D}_s, \label{chic0decay}\nonumber\\
  \mathcal{L}_{\psi D_sD_s} &=& ig_{\psi D_sD_s} \psi_\mu(\partial^\mu D_s^{\dag}D_s-D_s\partial^\mu D_s^{\dag}), \\
  \mathcal{L}_{D^\ast D_sK} &=& ig_{D^\ast D_sK} D^\ast _\mu(\partial^\mu D_s^\dag K-D_s\partial^\mu K^\dag).\nonumber
\end{eqnarray}

With the above preparation, we obtain the decay amplitude of the resonant contribution to the process $B^+\to D_s^+D_s^-K^+$, i.e.,
\begin{eqnarray}
 \mathcal{M}_{X} &=& \frac{g_{X}} {(p_2+p_3)^2-m_{X}^2 +im_{X}\Gamma_{X}}, \nonumber\\
 \mathcal{M}_{\psi} &=& \frac{g_{\psi} (P^\mu+p_1^\mu) (p_2^\nu-p_3^\nu) \tilde{g}_{\mu\nu}(\psi)} {(p_2+p_3)^2-m_{\psi}^2 +im_{\psi}\Gamma_{\psi}},\\
 \mathcal{M}_{D^\ast } &=& \frac{g_{D^\ast}(P^\mu+p_2^\mu) (p_1^\nu-p_3^\nu) \tilde{g}_{\mu\nu}(D^\ast )} {(p_1+p_3)^2-m_{D^\ast }^2 +im_{D^\ast }\Gamma_{D^\ast }}\nonumber
 \end{eqnarray}
with $\tilde{g}_{\mu\nu}(\psi)=-g_{\mu\nu} + (p_{2\mu}+p_{3\mu}) (p_{2\nu}+p_{3\nu}) /m_\psi^2$ and $\tilde{g}_{\mu\nu}(D^\ast )=-g_{\mu\nu} + (p_{1\mu}+p_{3\mu}) (p_{1\nu}+p_{3\nu})/m_{D^\ast }^2$, where the coupling constants are defined as $g_{X}=g_{XD_s\bar{D}_s}g_{BXK}$,  $g_{\psi}=g_{\psi D_sD_s}g_{B\psi K}$ and $g_{D^\ast}=g_{D^\ast D_sK}g_{BD^\ast D_s}$. 

In addition to the resonant contribution, there is the background contribution for the discussed process. In the realistic calculation, we consider a constant amplitude to simulate the background term, which is,
\begin{eqnarray}
\mathcal{M}_{bkg}=g_{bkg}.
\end{eqnarray}

And then, the total decay amplitude is the coherent sum of each resonant amplitudes and nonresonant background amplitude, i.e.,
\begin{eqnarray}\label{amptotal}
  \mathcal{M}_{Total} &=& \mathcal{M}_{bkg} + e^{i\phi_1}\mathcal{M}_{\chi_{c0}(2P)} + e^{i\phi_2}\mathcal{M}_{X_0(4140)} \nonumber\\
   &&+ e^{i\phi_3}\mathcal{M}_{\psi(4230)} + e^{i\phi_4}\mathcal{M}_{D^\ast (3^3S_1)},
\end{eqnarray}
where $\phi_i$ denotes the phase angle between the resonant amplitudes and the nonresonant background amplitude, which are considered as free parameters.  Moreover, the coupling constants $g_{bkg}$, $g_{\chi_{c0}(2P)}$, $g_{X_{0}(4140)}$, $g_{\psi(4230)}$ and $g_{D^\ast(3^3S_1)}$ are also treated as parameters, which can be determined by fitting the experimental data. While, the masses and widths of the involved intermediate states are fixed, and the concrete values of the resonant parameters are collected in Table \ref{Tab:input}.

The differential decay width of $B^+\to D_s^+D_s^-K^+$ relative to the invariant mass distribution of ${D_s^+D_s^-}$ is,
\begin{equation}
\label{dGamma}
 \frac{\mathrm{d}\Gamma}{\mathrm{d}m_{D_s^+D_s^-}}=\frac{1}{(2\pi)^5}\frac{1}{16m_B^2}\left|\mathcal{M}_{Total}\right|^2 \left| \vec{p}_1\right| \left| \vec{p}_2^\ast \right| \mathrm{d}\Omega_1 \mathrm{d}\Omega_{23}^\ast  ,
\end{equation}
where $\vec{p}_2^\ast $ and $\Omega_{23}^\ast $ are the three momentum and solid angle of the $D_s^+$ meson in the center of the $D_s^+D_s^-$ system mass frame, respectively.  Similarly, one can obtain the differential decay width relative to the invariant mass distributions of $D_s^+K^+$ and $D_s^-K^+$.

\begin{figure*}
  \centering
  \begin{tabular}{ccc}
\includegraphics[height=4.5cm]{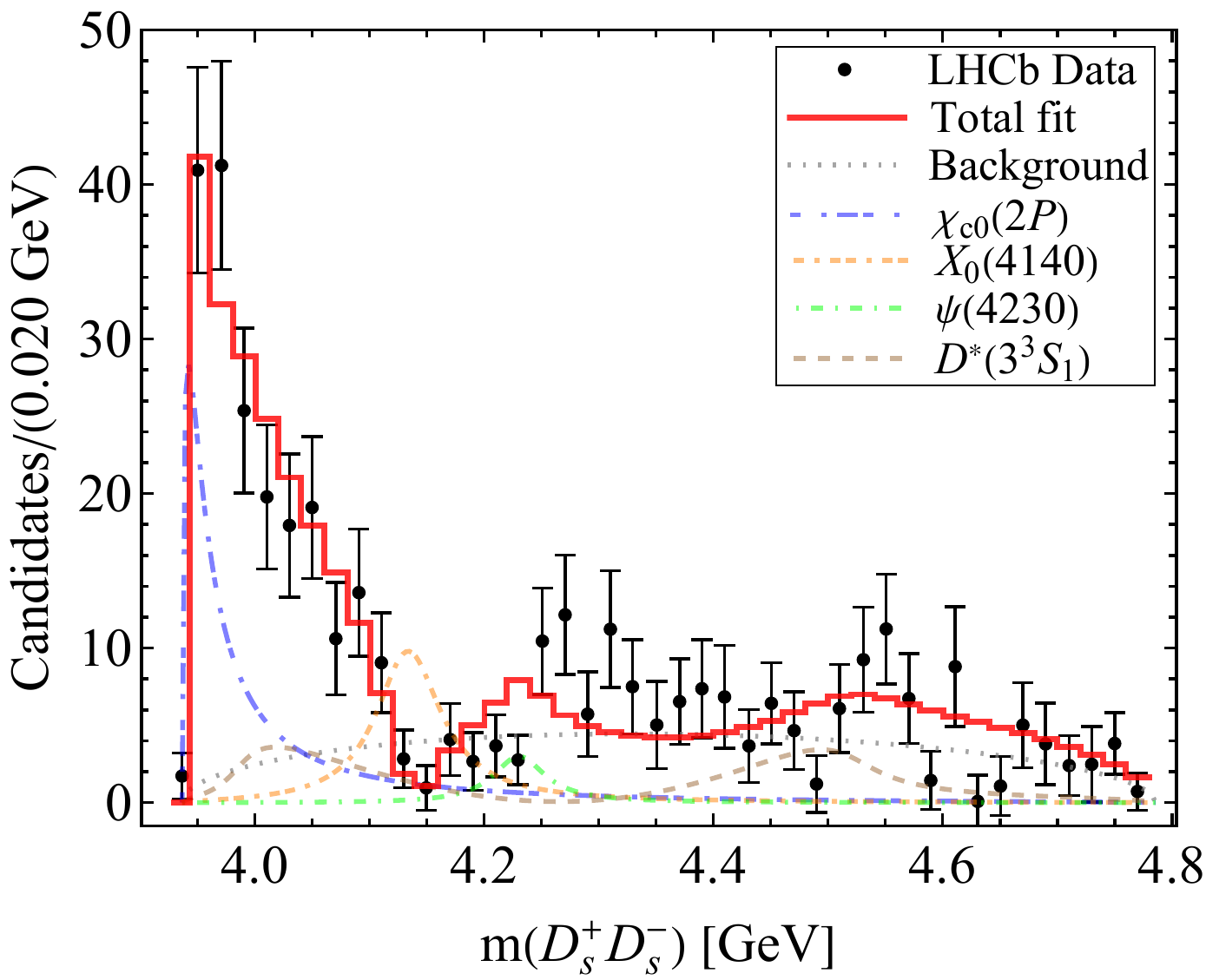}&
\includegraphics[height=4.5cm]{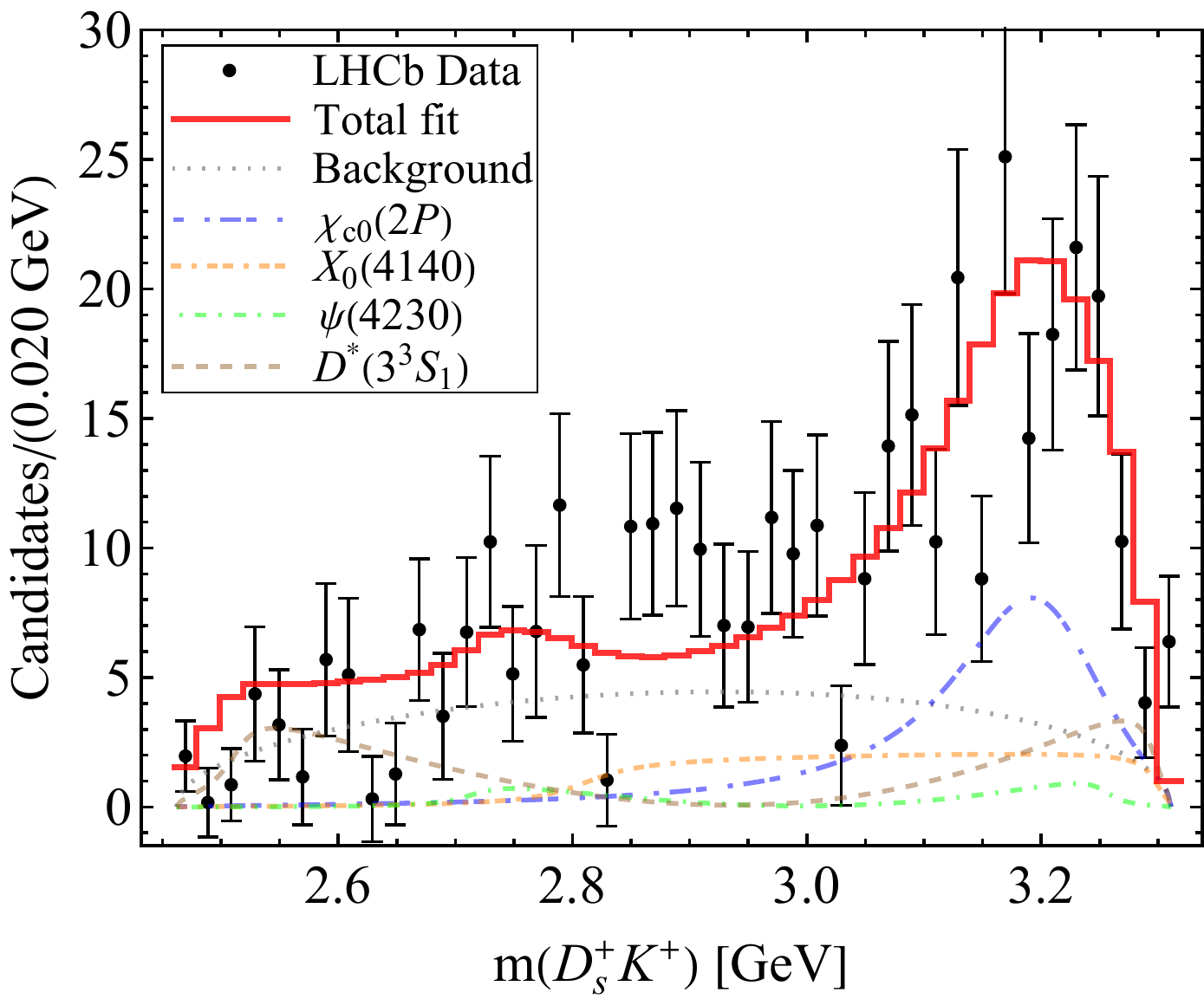}&
\includegraphics[height=4.5cm]{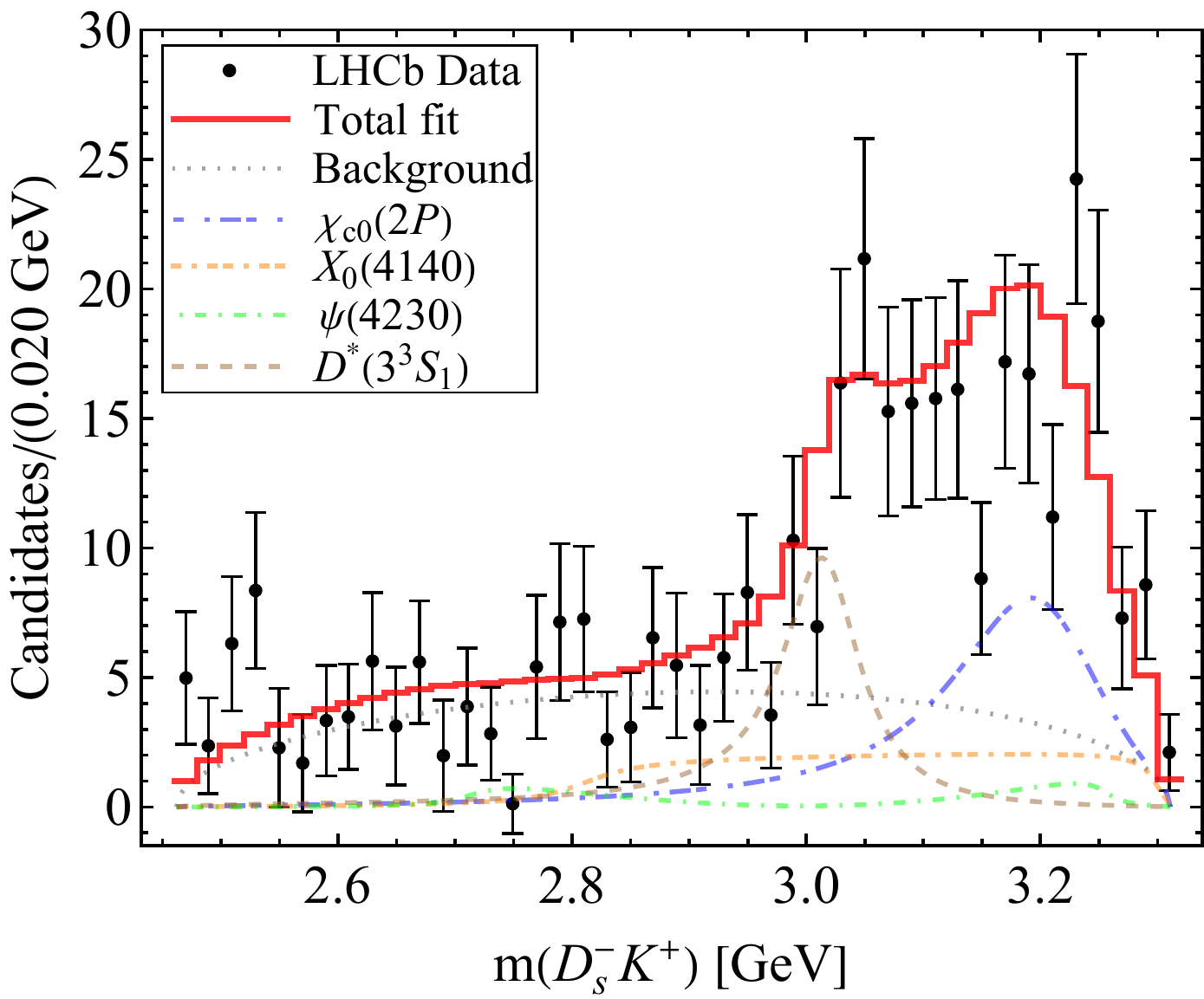}\\
(a)&(b)&(c)
  \end{tabular}
  \caption{(Color online.) Our best fit to the invariant mass spectra of (a) $D_s^+D_s^-$, (b) $D_s^+K^+$, and (c) $D_s^-K^+$ of the $B^+\to D_s^+D_s^-K^+$ process.  Here, the red solid step line indicates the total fitting results, while the individual contributions from background and the intermediate states are also given.}
  \label{invariantmass}
\end{figure*}

\begin{table}
\centering
 \renewcommand\arraystretch{1.25}
\caption{\label{Tab:input} The resonant parameters of the involved intermediate states.}
\begin{tabular}{p{3cm}<\centering p{2cm}<\centering p{2cm}<\centering}
	\toprule[1pt]\toprule[1pt]
	State & Mass (MeV) & Width (MeV)\\
	\midrule[1pt]
  $\chi_{c0}(2P)] $~\cite{LHCb:2020pxc}  & 3923.8 &  17.4 \\
  $X_0(4140)$~\cite{LHCb:2022vsv}            & 4133    & 67 \\
  $\psi(4230)$~\cite{LHCb:2022vsv}         & 4230 &  55 \\
  $D^\ast (3^3S_1) $~\cite{Song:2015fha}    & 3015 &  80.36 \\
  \bottomrule[1pt]\bottomrule[1pt]
\end{tabular}
\end{table}

\begin{table}[htb]
 \renewcommand\arraystretch{1.25}
  \centering
  \caption{\label{ParaTable} The values of the parameters obtained by fitting the LHCb data~\cite{LHCb:2022vsv}. }  
  \begin{tabular}{p{2cm}<\centering p{2.5cm}<\centering p{1.7cm}<\centering p{1.7cm}<\centering}
  \toprule[1pt]\toprule[1pt]
  Parameters         &Values ($\times10^{-7}$)          &Parameters       &Values\\
  \midrule[1pt]
  $g_{bkg}$          & $108.0\pm9.1$   & -- --           & -- -- \\
  $g_{\chi_{c0}(2P)}$  & $(73.8\pm2.9) \ \mathrm{GeV}^2$   &$\phi_1$  &1.01$\pm$0.01 \\
  $g_{X_0(4140)}$        & $(34.6\pm9.2) \ \mathrm{GeV}^2$  &$\phi_2$  &4.67$\pm$0.01 \\
  $g_{\psi(4230)}$       &8.0$\pm1.5$  &$\phi_3$  &1.12$\pm$0.03 \\
  $g_{D^\ast(3^3S_1)}$   &10.5$\pm1.0$  &$\phi_4$  &4.27$\pm$0.06 \\
    \bottomrule[1pt]\bottomrule[1pt]
  \end{tabular}
\end{table}

In the following, we can simultaneously reproduce the invariant mass distributions of $D_s^+ D_s^-$, $D_s^+ K^+$ and $D_s^-K^+$ of the process $B^+ \to K^+ D_s^+ D_s^-$. Our best fit to the experimental data is presented in Fig.~\ref{invariantmass} with $\chi^2/\mathrm{dof}=1.73$ and the corresponding parameter values are listed in Table~\ref{ParaTable}. Here, the candidates in Fig. \ref{invariantmass} can be directly associated to the differential width of $B^+ \to K^+ D_s^+ D_s^-$. Thus, our fitting parameters have been calibrated to reproduce the decay width of the process $B^+ \to K^+ D_s^+ D_s^-$ by using the experimental ratio $\mathcal{B}(B^+ \to K^+ D_s^+ D_s^-)/\mathcal{B}(B^+ \to K^+ D^+ D^-) =0.525 \pm 0.033$ \cite{LHCb2022Qian} and branching ratio $\mathcal{B}(B^+ \to K^+ D^+ D^-)=2.2\pm0.5\pm0.5$ \cite{Workman:2022ynf}. 
As shown in Fig. \ref{invariantmass} (a), these two structures reported by LHCb, i.e., the enhancement near the $D_s^+D_s^-$ threshold and the dip near 4.14 GeV, can be well reproduced by introducing the below-threshold $\chi_{c0}(2P)$ and the $X_{0}(4140)$, respectively. 
In addition, the event accumulation near 4.2 GeV can also be roughly described by the introduced $\psi(4230)$ contribution. 
It is obvious that the present work provides an alternative approach to explain the reported near threshold structure in the $D_s^+D_s^-$ invariant mass spectrum \cite{LHCb:2022vsv}. In our scenario, it is not necessary to introduce a new charmoniumlike state $X(3960)$ as treated by LHCb \cite{LHCb:2022vsv}.

In the present scheme, our fitting results also show a good consistency with the invariant mass distributions of $D_s^+K^+$ and $D_s^-K^+$. From Fig.~\ref{invariantmass} (b) and (c) one can notice that the reflection from the $\chi_{c0}(2P)$ plays the dominant role for resulting in the broad structure around 3.2 GeV in both $D_s^+K^+$ and $D_s^-K^+$ invariant mass spectra. However, as discussed in the last section, there exists an apparent difference between the $D_s^+ K^+$ and $D_s^- K^+$ invariant mass distributions. Thus, for the discussed $B^+\to D_s^+D_s^-K^+$ decay, we introduce an additional higher charmed meson besides the intermediate charmonia. After introducing the $D^\ast (3^3S_1)$ resonant contribution, the  shoulder-like shape in the $D_s^-K^+$ invariant mass distribution from 3.0 to 3.2 GeV can be well depicted. Moreover, the interference between the background term and the reflection contribution of the $D^\ast (3^3S_1)$ in the $D_s^+D_s^-$ invariant mass spectrum can naturally explain the bump near 4.5 GeV in the $D_s^+D_s^-$ invariant mass spectrum, which can also be an important hint of the intermediate $D^\ast (3^3S_1)$ contributing to the process $B^+\to D_s^+D_s^-K^+$. The refined properties of the unestablished charmed meson $D^\ast (3^3S_1)$ can be revealed by the more precise experimental measurement of the $B^+\to D_s^+D_s^-K^+$ decay at LHCb and Belle II.

\section{Explanation to the abnormal ratio involved in the $\chi_{c0}(2P)D^+D^-$ and $\chi_{c0}(2P)D_s^+D_s^-$ interactions\label{sec3}}

After reproducing the invariant mass distributions of $D_s^+ D_s^-$, $D_s^+ K^+$ and $D_s^- K^+$ in our scheme, we can further extract the fit fraction of $B^+\to K^+(\chi_{c0}(2P)\to D_s^+ D_s^-)$ to be $(20.8 \pm 11.0)\%$, where the systematical and statistical uncertainties of the experiments have been included. Then, one gets the ratio\footnote{In this work, we do not adopt the LHCb's value as given in Eq. (\ref{LHCbratio}). In our scheme, we extract
the ratio of ${\mathcal{B}[B^+\to D^+D^-K^+]\mathcal{FF}^{\chi_{c0}(2P)}_{B^+\to D^+D^-K^+}}$ and ${\mathcal{B}[B^+\to D_s^+D_s^-K^+]\mathcal{FF}^{\chi_{c0}(2P)}_{B^+\to D_s^+D_s^-K^+}}$ by the present experimental data.}
 \begin{eqnarray}
\frac{\mathcal{B}[B^+\to D^+D^-K^+]\mathcal{FF}^{\chi_{c0}(2P)}_{B^+\to D^+D^-K^+}}{\mathcal{B}[B^+\to D_s^+D_s^-K^+]\mathcal{FF}^{\chi_{c0}(2P)}_{B^+\to D_s^+D_s^-K^+}}=0.34\pm0.20.\label{h1}
 \end{eqnarray}
In fact, this value in Eq. (\ref{h1}) is the ratio of $\Gamma(\chi_{c0}(2P) \to D^+ D^-)$ and $\Gamma(\chi_{c0}(2P) \to D_s^+ D_s^-)$, which indicates the relative magnitude between the coupling of the $\chi_{c0}(2P)$ with the $D^+ D^-$ channel and that of the $\chi_{c0}(2P)$ with the $D_s^+ D_s^-$ channel.  Considering the phase space integral and taking the mass distributions of the $\chi_{c0}(2P)$ into account, the ratio of the coupling constants is obtained to be
\begin{eqnarray}\label{couplingratio}
  \frac{g_{\chi_{c0}(2P)D^+ D^-}} {g_{\chi_{c0}(2P)D_s^+ D_s^-}}=0.15\pm0.05. \label{Eq:Ratio}
\end{eqnarray}
If making a naive theoretical estimate and considering the SU(3) flavor symmetry, the ratio $g_{\chi_{c0}(2P)D^+ D^-}/g_{\chi_{c0}(2P)D_s^+ D_s^-}=1$ can be got. In fact, the $s$ quark has larger mass than the $u(q)$ quark, and the creation possibility of the $s\bar{s}$ quark pair from vacuum is less than that of the $u\bar{u}(d\bar{d})$ quark pair. Thus, this ratio is expected to be larger than 1. Obviously, the extracted ratio shown in Eq. (\ref{Eq:Ratio}) does not satisfy the above estimate. To some extent, there exists an unexpected reversal between coupling strengths  $g_{\chi_{c0}(2P)D^+D^-}$ and $g_{\chi_{c0}(2P)D_s^+D_s^-}$, which should be explained quantitatively.

As a main task of this work, we explain why the anomaly of this ratio can  happen by introducing the nontrivial node effect. Generally, the transition matrix of the open-charm process $\chi_{c0}(2P) \to D_{{(s)}}\bar{D}_{{(s)}}$ can be expressed as
 \begin{eqnarray}
 \mathcal{M}=\langle D_{{(s)}}\bar{D}_{{(s)}}|\mathcal{T}| \chi_{c0}(2P) \rangle, \label{Eq:TM}
 \end{eqnarray}
where $\mathcal{T}$ is the transition operator. 
In the quark pair creation (QPC) model \cite{Micu:1968mk,LeYaouanc:1972vsx,LeYaouanc:1973ldf,LeYaouanc:1974cvx,LeYaouanc:1977fsz,LeYaouanc:1977gm,
Guo:2019wpx,Guo:2022xqu,Song:2015nia}, which was extensively applied to study the Okubo-Zweig-Iizuka (OZI) allowed strong decays of conventional hadron in the past decades, the transition operator $\mathcal{T}$ is defined as
\begin{eqnarray}
\mathcal{T}&=&-3\gamma\sum_m\langle 1m;1-m|00\rangle
\int d{\mathbf k}_3d{\mathbf k}_4\delta^3({\mathbf k}_3+{\mathbf k}_4)
\nonumber\\&&\times\mathcal{Y}_{1m}\left(\frac{{\mathbf k}_3-{\mathbf k}_4}{2}\right)\chi_{1,-m}^{34}\phi_0^{34}\omega_0^{34}d_{3i}^\dag(\mathbf{k}_3)b_{4j}^\dag(\mathbf{k}_4) .
\end{eqnarray}
where $\mathcal{Y}_{1m}(\mathbf{k} )=|\mathbf{k}| Y_{1m}(\theta ,\phi)$, $\chi_{1,-m}^{34}$,  $\varphi_0^{34}= (u\bar{u}+d\bar{d}+s\bar{s})/\sqrt{3}$ and $\omega_0^{34} =\delta_{\alpha_3 \alpha_4}$ are the spatial, spin, flavor and color parts of the wave functions, respectively. $\alpha_3$ and $\alpha_4$ are the color indexes of the created quark pair from the vacuum. In the QPC model, the parameter $\gamma $ is introduced to represent the strength of the quark-antiquark pair creation from the vacuum, which satisfies the relation $\gamma_{u\bar{u}/d\bar{d}}=\sqrt{3}\gamma_{s\bar{s}}$ and is a universal value for the specific initial state system.

In the center-of-mass frame of charmonium $\chi_{c0}(2P)$, the transition matrix element in Eq. (\ref{Eq:TM}) is proportional to the overlap integral of the wave functions in the momentum space, which reads as
\begin{eqnarray}
&I(\mathbf{P},m_c,m_{\bar{c}},m_q)=\int d^3\mathbf{p} \; \Psi_{n_AL_AM_{L_A}}(\mathbf{P}+\mathbf{p}) \; \mathcal{Y}_{lm}(\mathbf{p}) \nonumber \\
& \quad\quad \times\Psi^\ast _{n_BL_BM_{L_B}}(\frac{m_q}{m_c+m_q}\mathbf{P}+\mathbf{p})
\Psi^\ast _{n_CL_CM_{L_C}}(\frac{m_q}{m_{\bar{c}}+m_q}\mathbf{P}+\mathbf{p}),\ \  \label{eq:integral}
\end{eqnarray}
where $\mathbf{P}$ denotes the momentum of either outgoing meson, and $m_{c(\bar{c})}$ and $m_q$ are masses of charm(anticharm) quark and the light quark, respectively. $\Psi_{nLM_L}(\mathbf{p})$ stands for the spatial part of the meson wave function, while the notation $A$, $B$ and $C$ refer to the $\chi_{c0}(2P)$, $D_{(s)}$ and $\bar{D}_{(s)}$ states, respectively. In the present study, the wave functions obtained in the unquenched potential model \cite{Wang:2019mhs} are employed. For the $n$th radial excitation of a meson system, its radial part of the spatial wave function $R_{nL}(p)$ has $(n-1)$ nodes.  For the involved $\chi_{c0}(2P)$ state, its radial spatial wave function only has one node, which makes us to easily distinguish positive and negative values of 
the radial wave function $R_{nL}^{\chi_{c0}(2P)}(p)$. 
Thus, the value of the integral in  Eq.~(\ref{eq:integral})  corresponding to some interval of integration can be either negative or positive, where these two parts can be partly cancelled with each other. In fact, such kind of cancellation is resulted from the node of radial wave function. The sensitivity of 
$I(\mathbf{P},m_c,m_{\bar{c}},m_q)$ to the position of the node of 
radial wave functions of the involved hadrons is called as node effect. 

By the effective Lagrangian of the  ${\chi_{c0}(2P)D\bar{D}}$ coupling, one can obtain the decay amplitude of $\chi_{c0}(3930)\to D\bar{D}$, which is $\mathcal{M}_{\chi_{c0}(2P) \to D\bar{D}} =g_{\chi_{c0}(2P) D\bar{D}}$. This decay amplitude can be calculated by the QPC model. 
By connecting the decay widths deduced from these two amplitudes, the coupling constants $g_{\chi_{c0}(2P) D\bar{D}}$ can be obtained. In the present work, only the parameter $\gamma$ in the QPC model should be fixed by reproducing the width of the $\chi_{c2}(2P)$, which is $34.2\pm 6.6\pm 1.1\; \mathrm{MeV}$~\cite{LHCb:2020pxc}. By the calculation, one gets $\gamma^2=40.9\pm 8.2$, which will be applied to the following investigation. 

\begin{figure}[htbp]
    \centering
 \includegraphics[width=0.45\textwidth]{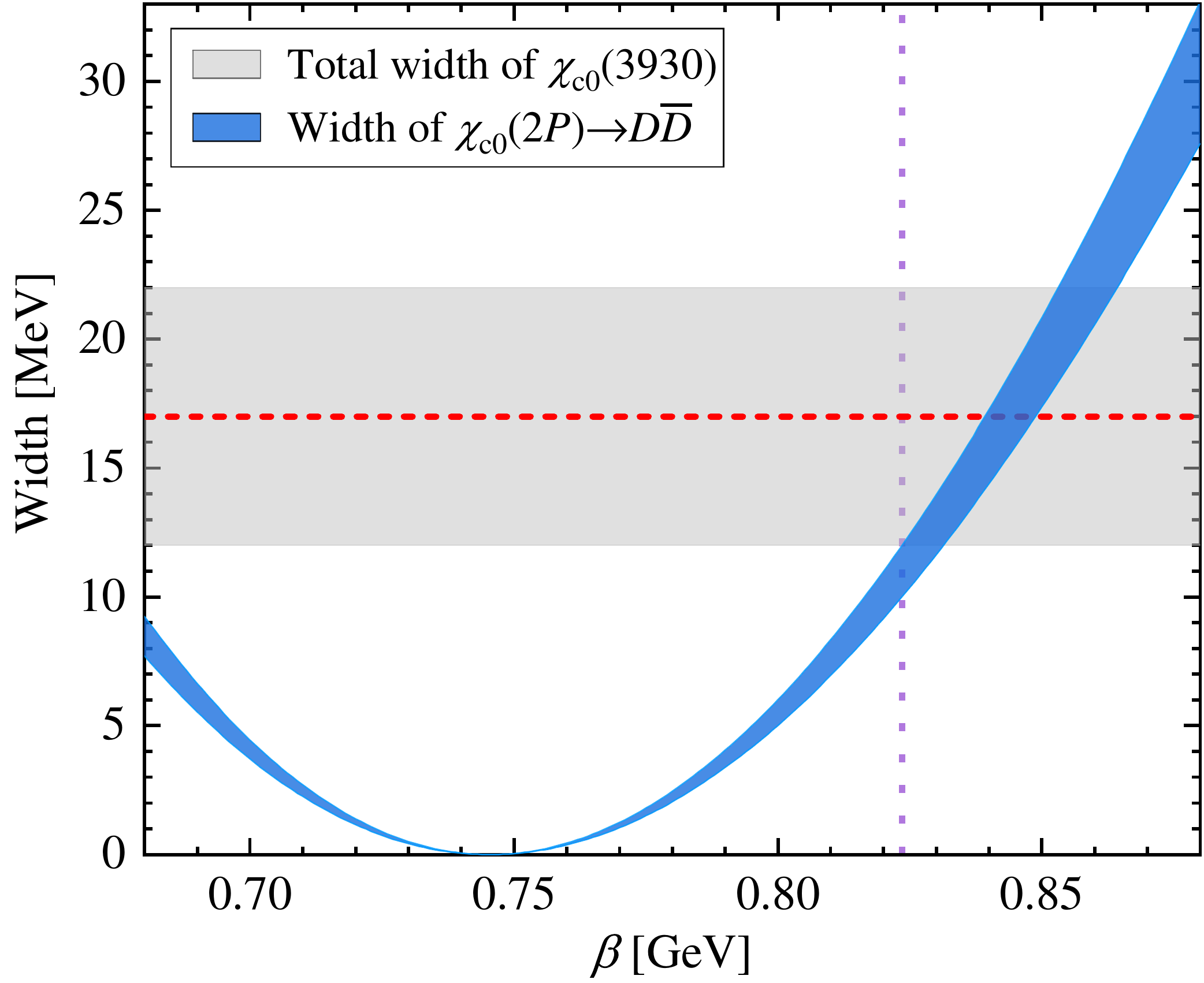}
    \caption{(Color online.) The decay width of $\chi_{c0}(2P)\to D\bar{D}$ dependent on the parameter $\beta$. Here, the blue band indicates the uncertainty resulted from the value of $\gamma$. The horizontal gray band with red dashed line stands for the measured width of the $\chi_{c0}(3930)$~\cite{LHCb:2020pxc}. The vertical dashed refers to $\beta=0.824$ GeV, where the estimated width of $\chi_{c0}(2P)\to D\bar{D}$ reaches up to the lower limit of the measured width of the $\chi_{c0}(3930)$. }
    \label{widthbeta}
\end{figure}

In order to show the node effect to the decay width and discuss possible theoretical uncertainty from the wave function, we adopt a simple harmonic oscillator (SHO) form to depict the radial wave function of the $\chi_{c0}(2P)$. Explicitly, it reads,
\begin{eqnarray}
 &&\Psi_{nlm}(\mathbf{p})=R_{nl}(p, \beta)Y_{lm}(\Omega_p), \\
 &&R_{nl}(p,\beta)=\frac{(-1)^n(-i)^l}{ \beta ^{3/2}}e^{-\frac{p^2}{2 \beta ^2}}\sqrt{\frac{2n!}{\Gamma(n+l+3/2)}}{\left(\frac{p}{\beta}\right)}^{l} \nonumber \\
 &&\quad\quad\quad\quad\times L_{n}^{l+1/2}\left(\frac{p^2}{ \beta ^2}\right),
\end{eqnarray}
where the node position of the radial wave function is determined by 
the oscillator parameter $\beta$. The value of $\beta$ for the $\chi_{c0}(3930)$ is 0.78 GeV, which is determined by the unquenched quark model~\cite{Wang:2019mhs}. Thus, in our calculation, we select a $\beta$ range to be $0.68\sim 0.88$ GeV
when taking the uncertainty of the theoretical model into account.

For the $\chi_{c0}(2P)$, $\chi_{c0}(2P) \to D\bar{D}$ is an OZI-allowed decay channel. Its  decay width dependent on the parameter $\beta$ is presented in Fig.~\ref{widthbeta}, where the node effect is obvious since the calculated decay width 
becomes smaller when the $\beta$ value is close to 0.746 GeV near the estimated value $\beta=0.78$ GeV
from the unquenched quark model~\cite{Wang:2019mhs}. 
This fact show that the $\chi_{c0}(2P) \to D\bar{D}$ decay can be suppressed by the node effect. Thus, it is possible to get weak coupling for the $\chi_{c0}(2P)D\bar{D}$ interaction when the node effect is considered. In Fig.~\ref{widthbeta}, we also list the experimental width of the $\chi_{c0}(3930)$ determined by LHCb
\cite{LHCb:2020pxc}. If reproducing this width value, we should take $\beta=(0.824\sim0.853)$ GeV, which is not obviously deviated from the value 0.78 GeV mentioned above. We have reason to believe that the node effect should be seriously considered when discussing the coupling of $\chi_{c0}(2P)D\bar{D}$.

\begin{figure}[htbp]
    \centering
    \includegraphics[width=0.48\textwidth]{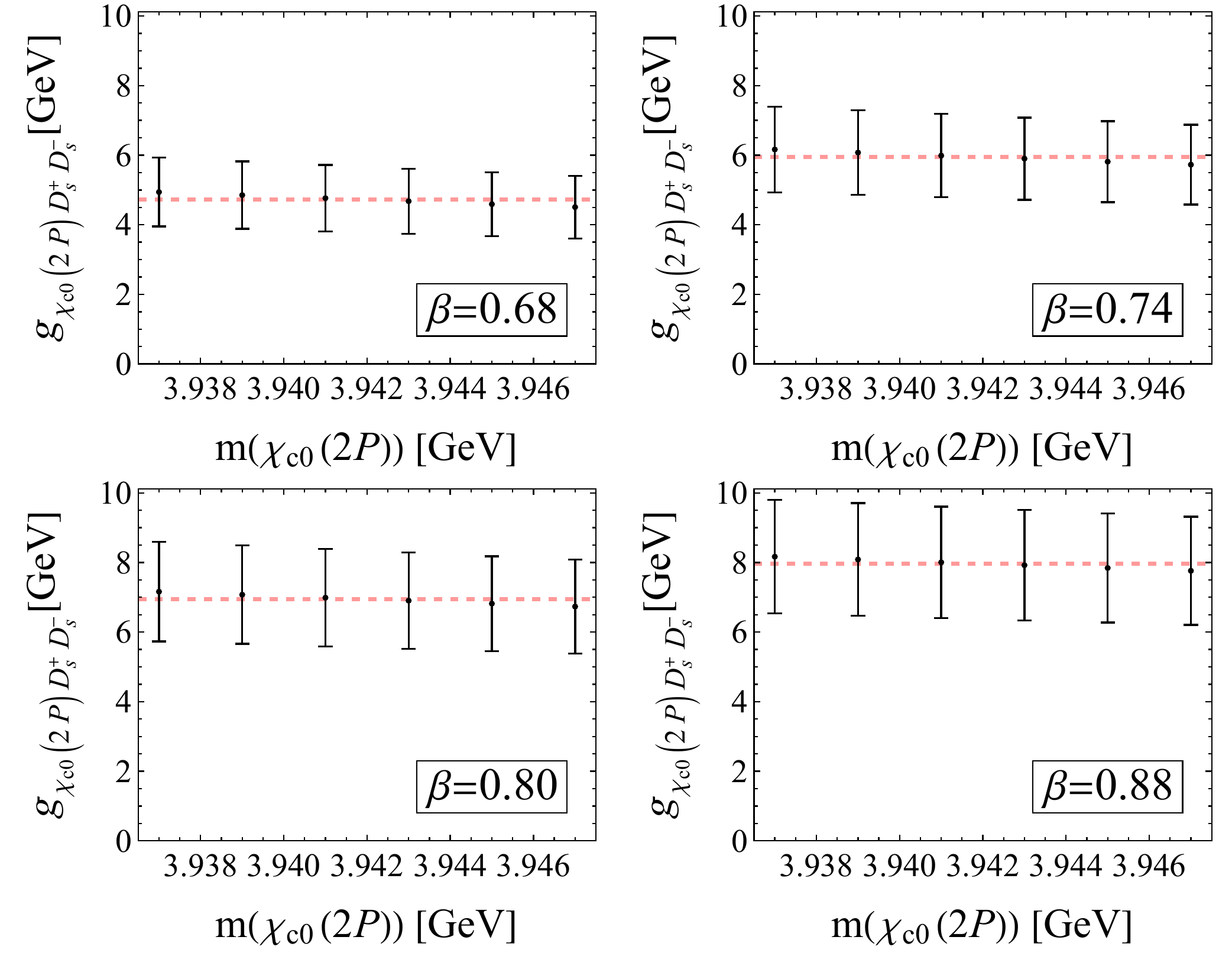}
    \caption{The mass dependence of the coupling $g_{\chi_{c0}(2P)D_s^+D_s^-}$ on different $\beta$ values. Here, six typical values of the mass of $\chi_{c0}(2P)$ are adopted, which are 3.937, 3.939, 3.941, 3.943, 3.945 and 3.947 GeV, respectively. The uncertainty of $g_{\chi_{c0}(2P)D_s^+D_s^-}$ results from the uncertainty of parameter $\gamma$, and the red dashed horizontal lines stands for the average values for different $\beta$.}
    \label{gDsDSbeta}
\end{figure}


In the following, we discuss 
how to estimate coupling constant $g_{\chi_{c0}(2P) D_s^+ D_s^-}$. If only taking central mass of the $\chi_{c0}(2P)$ as input, the process $\chi_{c0}(2P)\to D_s^+ D_s^-$ is kinematically forbidden. Although the $\chi_{c0}(2P)$ is below the threshold of $D_s^+ D_s^-$, the decay $\chi_{c0}(2P) \to D_s^+ D_s^-$  still can happen when considering the mass distribution of the $\chi_{c0}(2P)$\footnote{The mass gap  between the $\chi_{c0}(3930)\equiv\chi_{c0}(2P)$ and the $D_s^+ D_s^-$ threshold is $12\pm2$ MeV and the measured width of $\chi_{c0}(3930)$ is $17.4\pm 5.1\pm0.8$ MeV \cite{LHCb:2020pxc}.}, which is similar to the case of the $a_0(980)\to K\bar{K}$ decay. 
By the effective Lagrangian approach, the decay amplitude of the $\chi_{c0}\to D_s^+D_s^-$ is $\mathcal{M}_{\chi_{c0}(2P) \to D_s^+{D}_s^-} =g_{\chi_{c0}(2P) D_s^+{D}_s^-}$, by which the expression of this decay width can be obtained. For $\chi_{c0}\to D_s^+D_s^-$, its decay width can be also deduced by the QPC model. And then, the expression of $g_{\chi_{c0}(2P) D_s^+{D}_s^-}$ can be given by the above preparation with the connection of the decay widths from the effective Lagrangian approach and the QPC model. 

\begin{figure}[htbp]
    \centering
    \includegraphics[width=0.45\textwidth]{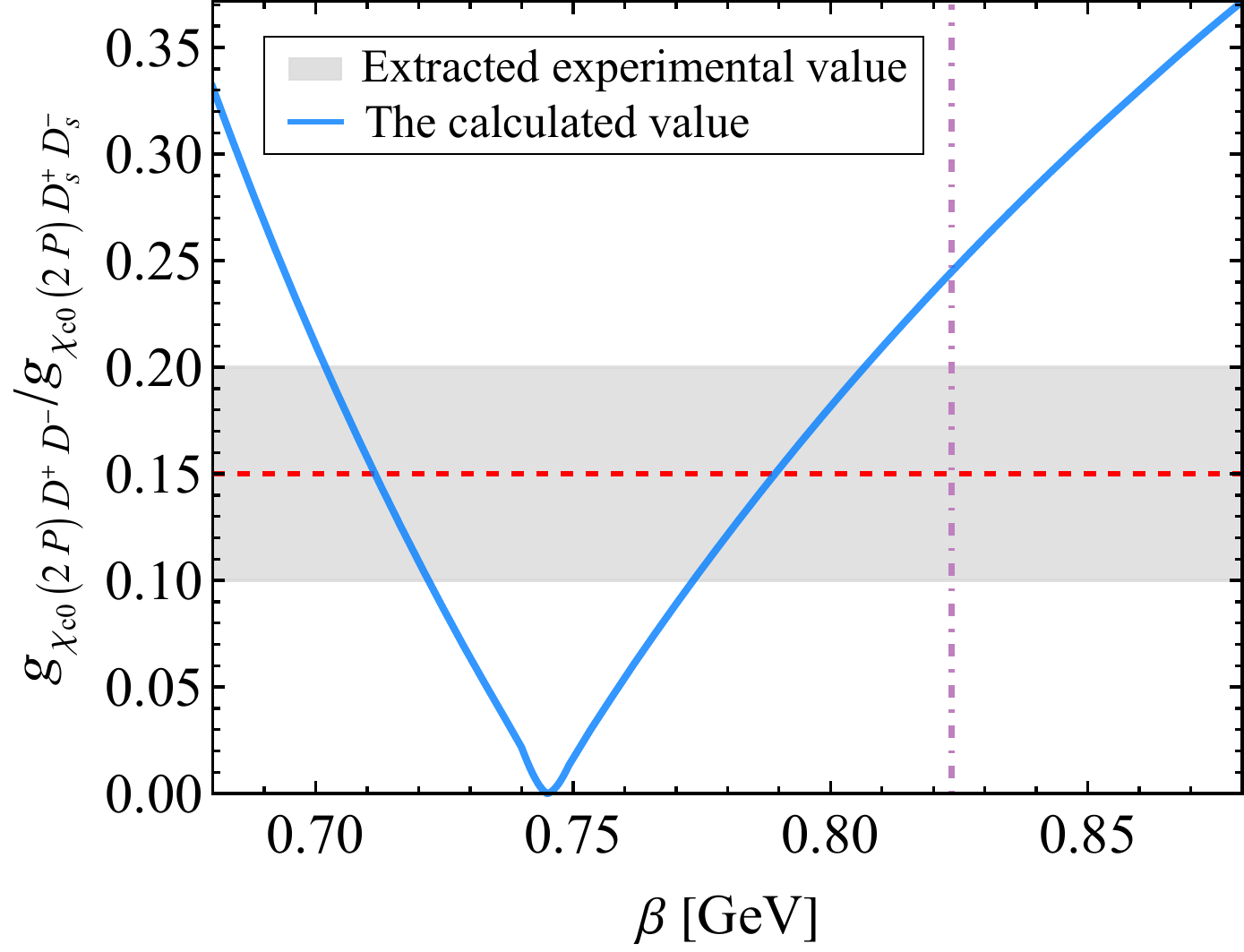}
    \caption{(Color online.) The $\beta$ dependence of the coupling constant ratio. The horizontal red dashed line with gray band indicates the coupling constant ratio extracted from LHCb data. The vertical dashed line corresponds to $\beta=0.824$ GeV. }
    \label{couplingbeta}
\end{figure}

We first check the $g_{\chi_{c0}(2P) D_s^+ D_s^-}$ value when taking six typical values of the mass distribution of the $\chi_{c0}(2P)$, all of which are above the $D_s^+ D_s^-$ threshold. By this way, these  
$g_{\chi_{c0}(2P) D_s^+ D_s^-}$ value corresponding to the above typical mass values are calculable. 
In Fig. \ref{gDsDSbeta}, the 
results of $g_{\chi_{c0}(2P)D_s^+D_s^-}$
dependent on several typical mass values of the $\chi_{c0}(2P)$ are collected, where four typical $\beta$ values are taken. We may conclude that the obtained $g_{\chi_{c0}(2P)D_s^+D_s^-}$ values are almost stable when changing the mass of the $\chi_{c0}(2P)$ as shown in Fig. \ref{gDsDSbeta}. 
Thus, we may adopt an extrapolation
to estimate the realistic $g_{\chi_{c0}(2P)D_s^+D_s^-}$ value, which is considered to be consistent with the obtained stable value of $g_{\chi_{c0}(2P)D_s^+D_s^-}$ corresponding to the mass range above the
$D_s^+ D_s^-$ threshold. 
We also check the node effect of $g_{\chi_{c0}(2P)D_s^+D_s^-}$. Our result shows that the coupling $g_{\chi_{c0}(2P)D_s^+D_s^-}$ is not sensitive to these selected $\beta$ values. 
For the given $\beta$ range\footnote{The node effect of the coupling $g_{\chi_{c0}(2P)D_s^+ D_s^-}$ is obvious if taking $\beta<0.5$ GeV, which is deviated from the $\beta$ value fixed by the unquenched quark model \cite{Wang:2019mhs}.}, the node effect to $g_{\chi_{c0}(2P)D_s^+ D_s^-}$ is not apparent, which is completely different from the behavior of the node effect to $g_{\chi_{c0}(2P)D^+D^-}$.

With the above preparation, we further present the ratio of $g_{\chi(c0)(2P) D_s^+D_s^-}$ and $g_{\chi_{c0}(2P)D_s^+ D_s^-}$ in Fig.~\ref{couplingbeta}, which is dependent on the parameter $\beta$. 
For a comparison, we also list the corresponding ratio extracted from LHCb data (see the value in Eq.~(\ref{Eq:Ratio})). 
Fig.~\ref{couplingbeta} illustrates how the node effect plays the crucial role to be responsible for explaining the coupling reversal phenomenon. Here, in an extreme case, the coupling constants ratio $g_{\chi_{c0}(2P)D^+D^-}/g_{\chi_{c0}(2P)D_s^+D_s^-}$ could be zero when $\beta=0.746$ GeV. 
In particular, we find the obtained ratio can overlap with the one extracted from LHCb data if the $\beta$ range is $(0.702-0.722) \ \mathrm{GeV}$ or $(0.773-0.807) \ \mathrm{GeV}$, where the later one is a litter bit smaller than the $\beta$ parameter range ($(0.824-0.853) \ \mathrm{GeV}$) determined by the width of the $\chi_{c0}(3930)$. 
This small discrepancy of the $\beta$ range can be understood. 
Since the width of the $\chi_{c0}(3930)$ is composed of its partial decay widths from open-charm decay channel, two gluon $gg$ process, the hidden-charm channel like $\omega J/\psi$ \cite{Belle:2009and}, and even electromagnetic transition, the realistic $\beta$ range becomes smaller if adopting the $\chi_{c0}(2P)\to D\bar{D}$ partial decay width to determine it. 
Finally, the anomaly of the ratio $g_{\chi_{c0}(2P)D^+D^-}/g_{\chi_{c0}(2P)D_s^+D_s^-}$ can be explained well, which not only reflects the importance of the node effect, but also enforces our scenario that the near $D_s^+D_s^-$ threshold enhancement can be due to the $\chi_{c0}(2P)$ contribution.


\section{Summary}\label{sec4}

Very recently, the LHCb Collaboration announced a near threshold enhancement, referred to be the $X(3960)$ before deciphering its nature, in the $D_s^+D_s^-$ invariant mass distribution of the process $B^+ \to D_s^+D_s^-K^+$~\cite{LHCb:2022vsv}. The $X(3960)$ favors $J^{PC}=0^{++}$ suggested by LHCb~\cite{LHCb:2022vsv}. If relating the $X(3960)$ to the $\chi_{c0}(3930)$ reported in the $B^+\to D^+D^-K^+$ decay ~\cite{LHCb:2020pxc}, LHCb obtained the ratio $\Gamma(X\to D^+D^-) / \Gamma(X\to D_s^+D_s^-)$ to be $0.29\pm0.09\pm0.10\pm0.08 $~\cite{LHCb:2022vsv}, which is not consistent with of the naive expectation of assigning the $X(3960)$ as a charmonium $\chi_{c0}(2P)$. Thus, LHCb claimed that the observed $X(3960)$ should be a candidate of the $c\bar{c}s\bar{s}$ tetraquark~\cite{LHCb:2022vsv}. 

In this work, we propose that this enhancement phenomenon near the $D_s^+D_s^-$ threshold is resulted from a conventional $P$-wave charmonium $\chi_{c0}(2P)$ below the $D_s^+D_s^-$ threshold  \cite{Duan:2020tsx,Duan:2021bna,Liu:2009fe,Belle:2009and,BaBar:2012nxg,LHCb:2020pxc,Chen:2012wy}. 
For testing such scenario, a combined fit to the measured $D_s^+D_s^-$, $D_s^+K^+$, and $D_s^-K^+$ invariant mass spectra is performed. Here, by introducing the $\chi_{c0}(2P)$ contribution, indeed we reproduce the near threshold enhancement in the $D_s^+D_s^-$ invariant mass spectrum well. 
For depicting the whole $D_s^+D_s^-$ invariant mass spectrum given by LHCb, we also introduce
higher charmonium $\psi(4230)$ \cite{BESIII:2014rja,BESIII:2016bnd,BESIII:2016adj,Wang:2019mhs,Wang:2017sxq,Qian:2021neg,Chen:2014sra,Chen:2017uof} and a scalar charmoniumlike state $X_0(4140)$, which are as the intermediate states in the discussed $B^+ \to D_s^+D_s^-K^+$ decay. 
Additionally, we also reveal that 
the predicted charmed meson $D^*(3^3S_1)$ with mass around 3015 MeV \cite{Song:2015fha,Wang:2016krl} may play crucial role when describing the $D_s^-K^+$ invariant mass spectrum well. 

In this work, another important issue is to explain the anomaly of the ratio of $\Gamma(X\to D^+D^-)/\Gamma(X\to D_s^+D_s^-)$ indicated by LHCb. Our study reveals that 
the node effect of the spatial wave function of the $\chi_{c0}(2P)$ is the main reason to result in this anomaly of the ratio. Finally, explaining 
the anomaly of the ratio of $\Gamma(X\to D^+D^-)/\Gamma(X\to D_s^+D_s^-)$ enforces our scenario again, i.e., the newly observed enhancement structure near the $D_s^+D_s^-$ threshold can be due to the $\chi_{c0}(2P)$ contribution. 

In a summary, after observing the  enhancement phenomenon near the 
$D_s^+D_s^-$ threshold in $B^+ \to D_s^+D_s^-K^+$ \cite{LHCb:2022vsv}
inspired interest of the community in deciphering its underlying mechanism. LHCb tried to assign 
this observed enhancement structure as a tetraquark  \cite{LHCb:2022vsv}. In fact, there is no hurry to introduce exotic state assignment before exhausting possible conventional explanations. 
In this work, an alternative explanation is proposed, where the observed near threshold enhancement structure and the corresponding anomaly of the ratio of $\Gamma(X\to D^+D^-)/\Gamma(X\to D_s^+D_s^-)$ indicate 
the existence of a $\chi_{c0}(2P)$ charmonium below the $D_s^+D_s^-$ threshold, which is a crucial step to establish 
the $\chi_{c0}(2P)$ charmonium. 

\section*{Acknowledgements}
This work is supported by the China National Funds for Distinguished Young Scientists under Grant No. 11825503 and  No. 12175037, National Key Research and Development Program of China under Contract No. 2020YFA0406400, and the 111 Project under Grant No. B20063, the National Natural Science Foundation of China under Grant No. 12047501. J.-Z.W. is also supported by the National Postdoctoral Program for Innovative Talent.

\end{document}